\documentclass[useAMS,usenatbib]{mn2e}
\usepackage{graphicx}
\usepackage{amssymb}
\usepackage{natbib}
\newcommand{\msun}{{M$_\odot$}}

\title{H$_2$ Suppression with Shocking Inflows: Testing a Pathway for Supermassive Black Hole Formation}

\author[R. Fernandez et al.]{Ricardo Fernandez$^{1}$, Greg L. Bryan$^{1}$, Zoltan Haiman$^{1}$, and Miao Li$^{1}$\\
$^{1}$Department of Astronomy, Columbia University, 550 West 120th Street, New York, NY 10027, USA}

\begin{document}

\date{}

\pagerange{\pageref{firstpage}--\pageref{lastpage}} \pubyear{2013}

\maketitle

\label{firstpage}

\begin{abstract}
The presence of quasars at redshifts $z > 6$ indicates the existence of supermassive black holes (SMBHs) as massive as a few times $10^9$ M$_{\odot}$, challenging models for SMBH formation.  One pathway is through the direct collapse of gas in $T_{\rm vir} \gtrsim 10^4$ K halos; however, this requires the suppression of H$_2$ cooling to prevent fragmentation.  In this paper, we examine a proposed new mechanism for this suppression which relies on cold-mode accretion flows leading to shocks at high densities ($n > 10^4$ cm$^{-3}$) and temperatures ($T > 10^4$ K).  In such gas, H$_2$ is efficiently collisionally dissociated.  We use high-resolution numerical simulations to test this idea, demonstrating that such halos typically have lower temperature progenitors, in which cooling is efficient.  Those halos do show filamentary flows; however, the gas shocks at or near the virial radius (at low densities), thus preventing the proposed collisional mechanism from operating.  We do find that, if we artificially suppress H$_2$ formation with a high UV background, so as to allow gas in the halo center to enter the high-temperature, high-density ``zone of no return", it will remain there even if the UV flux is turned off, collapsing to high density at high temperature.  Due to computational limitations, we simulated only three halos.  However, we demonstrate, using Monte Carlo calculations of $10^6$ halo merger histories, that a few rare halos could assemble rapidly enough to avoid efficient H$_2$ cooling in all of their progenitor halos, provided that the UV background exceeds $J_{21}\sim$ few at redshifts as high as $z\sim 20$. \end{abstract}



\section{Introduction}

Dynamical evidence indicates that most nearby galaxies harbor a central supermassive black hole \citep[e.g.,][]{Ferrarese2005}, including our own Milky Way, which hosts a central SMBH with mass $\thicksim 10^6$ M$_{\odot}$ \citep{Ghez2005}. Furthermore, the discovery of quasars at redshifts greater than 6 signals the existence of SMBHs as massive as a few times $10^9$ M$_{\odot}$ at an epoch when the Universe was less than a billion years old \citep[e.g.,][]{Fan2006, Mortlock2011}. Such massive and early SMBHs pose a challenge to current models of their formation.

One possible formation scenario is the growth of a remnant black hole (BH) seed, generated from a population III star ($\thicksim 100$ M$_{\odot}$), by mergers and gas accretion \citep[e.g.,][]{Haiman2001, Volonteri2003, Li2007}.  However, this formation scenario poses certain difficulties. The time to assemble a $10^9$ M$_{\odot}$ SMBH by standard Eddington accretion is comparable to the age of the universe at $z \thicksim 6$ and it is unlikely that the seed BH will have continual accretion due to negative feedback and merger-induced gravitational recoils \citep{Alvarez2009, Milosavljevic2009, Tanaka2009, Jeon2011, Tanaka2012}. 

An alternative pathway is the direct collapse of metal-free primordial gas with virial temperature $\gtrsim 10^4$ K into a BH seed of mass $10^4-10^6$ M$_{\odot}$ \citep{Oh2002, Bromm2003, Regan2009a, Shang2010}. Such a large seed requires many fewer Salpeter times to grow to quasar size and so bypasses many of the difficulties of Eddington growth of stellar BH seeds.  The exact mechanism by which the collapse occurs is not entirely clear \citep[e.g.,][]{Bromm2003, Begelman2008, Begelman2010}; however, a vital condition for this scenario is that the collapsing gas avoids fragmentation into stars.  A natural way to avoid fragmentation is to have a long Jeans length due to a high gas temperature.  The temperature of the gas depends on the interplay of atomic and molecular cooling. In the absence of H$_2$, a halo with $T_{\rm vir} \gtrsim 10^4$ K cools to $\thicksim8000$ K by atomic hydrogen. Including H$_2$, the halo cools further to $\thicksim 200$ K. The corresponding Jeans mass at characteristic central densities $M_J \approx 10^6$ M$_{\odot} (T/10^4$ K $)^{3/2}$ is $10^3$ M$_{\odot}$ for the latter, suggesting the formation of a Pop III star, and $10^6$ M$_{\odot}$ for the former, suggesting direct collapse into a massive BH.  Therefore, a necessary condition for direct collapse is to prevent cooling by H$_2$.

The suppression of H$_2$ can be accomplished by a strong far ultraviolet (UV) radiation flux in the Lyman-Werner (LW) bands, $J_{21} \gtrsim 10^2-10^3$ (in units of $10^{-21}$ erg s$^{-1}$ cm$^{-2}$ Hz$^{-1}$ sr$^{-1}$)  \citep{Omukai2001, Bromm2003, Shang2010}. This photo-dissociates the molecular hydrogen.  However, only a small subset $\lesssim 10^{-6}$ of all atomic cooling halos are estimated to be exposed to such levels, due to the presence of close luminous neighbor \citep{Dijkstra2008, Agarwal2012}.  
This makes it difficult to explain the production of the observed quasar population at $z > 6$.  
Note that the threshold only increases in the presence of a cosmic-ray/X-ray flux \citep{Inayoshi2011}.

\citet{Inayoshi2012} have proposed an alternative mechanism for the suppression of H$_2$ cooling that does not depend on having a high UV flux. In this scenario, cold accretion flows penetrate to the center of the halo, colliding with each other and shocking to produce hot and dense gas. The post-shock layer cools efficiently due to atomic hydrogen cooling and contracts isobarically until the gas reaches $\thicksim8000$ K.  If the shocked gas at this high temperature is already at a high enough number density $n \gtrsim 10^4$ cm$^{-3}$, then H$_2$ rotational-vibrational levels reach local thermodynamic equilibrium, and collisional dissociation can destroy the molecular hydrogen. Crucially, once the gas is shocked to this high-temperature, high-density regime, it will no longer be able to cool via ${\rm H_2}$, even in the absence of any LW radiation. Throughout this paper, we will thus refer to this regime (defined more precisely below) as the ``zone of no return''. \citet{Inayoshi2012} have argued that this mechanism may be able to produce a massive BH seed without strong radiative feedback; however, their numerical experiments focused on one-zone models and so questions still remain about the applicability of this pathway in cosmological simulations.  In particular, only fluid elements that have a sufficiently high temperature and density have their fragmentation suppressed, and it is not clear if: (1) the gas which ends up in halos that with such large virial temperatures is not first processed through lower mass halos in which fragmentation and star formation can occur\footnote{Although we note that \citet{Inayoshi2012} have argued that a small amount of metals does not change their results.}, and (2) if a sufficient amount of gas enters the ``zone of no return'' for this mechanism to be important in realistic halos.

To address these questions, we explore the possibility of H$_2$ suppression via cold accretion shocks by conducting numerical simulations. This paper is organized in the following manner. In Section~\ref{sec:numerical} we describe the ingredients and initial setup of the code. In Section~\ref{sec:results} we describe the results of our numerical simulations followed by a discussion in Section~\ref{sec:discussion}. Finally, in Section 5 we summarize our conclusions.


\section{Numerical Method}
\label{sec:numerical}

\begin{table}
       \begin{center}
	\begin{tabular}{@{}ccccc}
	\hline
	Halo & $z_{\mathrm{col}}$ & $T_{\mathrm{vir, col}} $ (K) & $M_{\mathrm{vir, col}} $ (M$_{\odot}$) &  $r_{\mathrm{vir, col}}$ (pc) \\ 
	\hline
	A & 17 & 7.80$ \times 10^3$ & 1.76$ \times 10^7$ & 462.99 \\
        B & 13 & 1.05$ \times 10^4$ & 3.98$ \times 10^7$ & 778.78 \\
        C & 12 & 8.50$ \times 10^3$ & 3.23$ \times 10^7$ & 710.25 \\ 
       \hline
	\end{tabular}
	\caption{Virial quantities for the three halos selected for resimulation from the low resolution run. The maximum level of refinement was set to 4 and radiative cooling was turned off.}
	\label{low-res-run}
	\end{center}
\end{table}

The simulations were performed with the publicly available Eulerian adaptive mesh refinement (AMR) Enzo code. \citep{Bryan1999, Norman1999, OShea2004, Bryan2013}. The code implements an N-body particle mesh technique \citep{Efstathiou1985, Hockney1988} to follow the dynamics of the dark matter particles and an Eulerian AMR method \citep{Berger1989} for the gas. In addition, Enzo provides modules which compute the radiative cooling of the gas as well as solve the chemical reaction network of a primordial mixture of H and He. Our simulations use the H$_2$ cooling function of \cite{Galli1998} and solve the non-equilibrium evolution of the following nine species: H, H$^+$, He, He$^+$, He$^{++}$, H$^-$, H$_2$, H$^+_2$, and e$^{-}$ \citep{Abel1997, Abel2000}.  
Density-dependent collisional dissociation \citep{Martin1996} is important -- we include this
with a rate as described in~\citet{Shang2010}.

We use a set of zoom simulations to focus on a number of halos selected from a 1 $h^{-1}$ Mpc comoving box, with a root grid resolution of $128^3$, using standard $\Lambda$ cold dark matter model parameters: $\Omega_{\Lambda,0} = 0.721$, $\Omega_{m,0} = 0.233$, $\Omega_b = 0.233$, $\sigma_8 = 0.817$, $n_s = 0.96$ and $h=0.701$ \citep{Komatsu2009}. Initially we performed a low resolution run with the maximum refinement level set to 4 and radiative cooling turned off to inhibit the gas from collapsing to high densities.  We evolved this simulation from $z=99$ to $z=10$. Then we applied the HOP halo finder \citep{Hop} to the resulting data files at various redshifts to identify halos with masses corresponding to virial temperatures $\gtrsim 10^4$K. Throughout this paper, we adopt the relation $T_{\rm vir}= 0.75\times1800 (M/10^6{\rm M_\odot})^{2/3} (1+z)/21$ K between halo mass and virial temperature.  This is consistent with the commonly adopted version for neutral primordial gas with mean molecular weight $\mu=1.2$ \citep{BryanNorman1998}, except that we reduced the normalization by a factor of 0.75.  We have found that this correction agrees better with our simulations - that is, it yields $T_{\rm vir}=10^4$K for halos at the redshift and mass when they begin to cool efficiently via atomic H.

\begin{figure*}
\centerline{\includegraphics[scale=0.6]{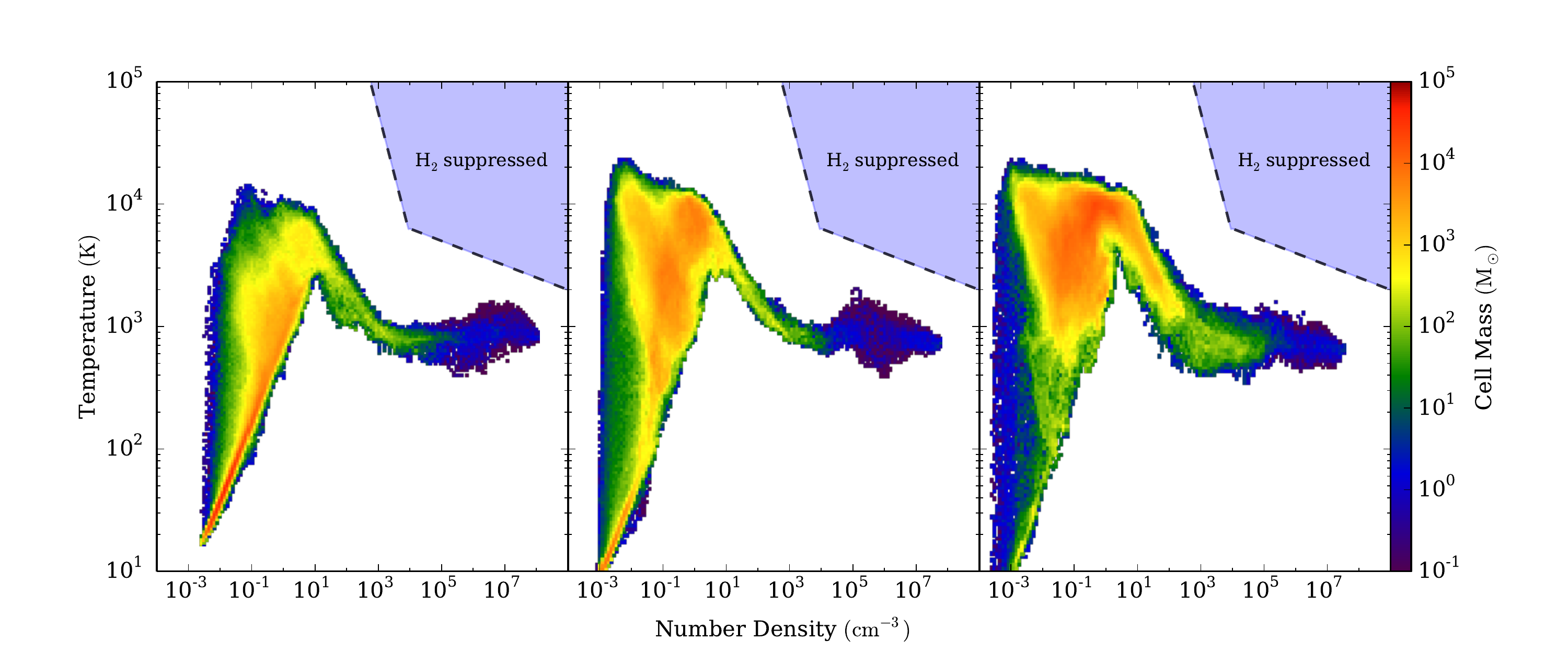} }
\caption{Phase plots of number density and temperature for each of the $J_{21} = 10$ simulations (from left to right: halos A, B and C).  The color indicates how much mass is at each point in the phase-diagram.  In all three simulations, no cells were shocked heated to the ``zone of no return'', shown by the dashed line and defined as in \citet{Inayoshi2012}.}
\label{phase-plots}
\end{figure*}

Three halos where selected at random (see Table~\ref{low-res-run}) to be re-run at high-resolution.  
We regenerated the initial conditions for the volumes, adding three nested grids that enclosed the Lagrangian volume of each halo.  Since each additional grid doubles the spatial resolution, this resulted in an innermost grid with an effective resolution of $1024^3$ and a dark matter particle mass of $\thicksim 85$ M$_{\odot}$.  Radiative cooling and multi-species were turned on to self-consistency follow the build-up of molecular hydrogen.  During the course of the simulation each cell was adaptively refined using the following three criteria: baryon mass, dark matter mass and Jeans length. For the first two criteria, refinement is added when the baryon or dark matter mass exceeds four times the mass of the initial most refined cell, corresponding to mass resolutions of 68 and 340 M$_\odot$ for the baryons and dark matter, respectively.  The third criterion enforces the \cite{Truelove1997} condition which states that at least four cells should resolve the Jeans length to avoid artificial fragmentation. In our simulations the Jeans length was resolved by sixteen cells to be sure that we adequately followed the collapse.  Based on these criteria, the simulations were allowed to refine to a maximum level of 18, which corresponds to a comoving scale of 0.0298 $h^{-1}$ pc.   The dark matter distribution was smoothed at refinement level 13 (about 0.065 proper pc at $z=20$) to suppress numerical effects from the discreteness of DM particles.  

We carried out two sets of runs for each of the halos, which differed only in the background LW flux that we adopted.  In the first set of runs, we used $J_{21} = 10$, where $J_{21}$ is the specific intensity in the Lyman-Werner bands (11.2-13.6 eV) in units of $10^{-21}$ erg cm$^{-2}$ sr$^{-1}$ Hz$^{-1}$.  This corresponds to a typical (but slightly high) value in the late pre-ionization period \citep{Dijkstra2008} and is well below that required to suppress H$_2$ radiatively \citep[e.g.,][]{Shang2010, Wolcott-Green2011}.  We use these simulations to determine if the H$_2$ suppression mechanism suggested by \citet{Inayoshi2012} can be responsible for halting fragmentation in these halos.  We carry out these runs until they collapse to high densities and then examine the resulting gas distribution.  As we will show, these halos do form an abundant supply of H$_2$, and so the \citet{Inayoshi2012} mechanism by itself does not appear to be sufficient to allow direct collapse.  In fact, cooling and collapse set in well before the virial temperature reaches $10^4$ K.

In a second set of simulations, we adopt a much higher value of the LW background, in particular we take $J_{21} = 10^5$, which is well above the critical flux required to suppress H$_2$ formation and cooling. We evolve these simulations until their virial temperatures are above $10^4$ K, which allows us to (artificially) run the halo until it has a virial temperature sufficiently high for the \citet{Inayoshi2012} mechanism to operate.  We then turn the LW background down to $J_{21} = 10$ to see if and how the purely collisional H$_2$ suppression acts.  These runs are intended as a academic exercise to see if gas that is in or near the ``zone of no return''  will indeed stay there.


\section{Results}
\label{sec:results}

In the following sections, we present the results of our numerical simulations, first focusing on runs with low background flux, and then follow with high-UV background simulations.

\subsection{${J_{21} = 10}$ Runs}

\begin{table}
	\begin{tabular}{@{}ccccc}
	\hline
	Halo & $z_{\mathrm{col}}$ & $T_{\mathrm{vir, col}} $ (K) & $M_{\mathrm{vir, col}} $ (M$_{\odot}$) &  $r_{\mathrm{vir, col}}$ (pc) \\ 
	\hline
          A & 21.81 & 2.86$ \times 10^3$ & 2.72$ \times 10^6$ & 199.14 \\
          B & 16.12 & 6.43$ \times 10^3$ & 1.41$ \times 10^7$ & 421.10 \\
          C & 12.89 & 0.81$ \times 10^4$ & 2.75$ \times 10^7$ & 642.13 \\ 
          \hline
	\end{tabular}
	\caption{Virial quantities of our three halos, as defined at the indicated collapse redshifts $z_{\rm col}$ for the $J_{21} = 10$ runs. }
	\label{high-res-run}
\end{table}

\begin{figure*}
\begin{center}
\includegraphics[scale=0.9]{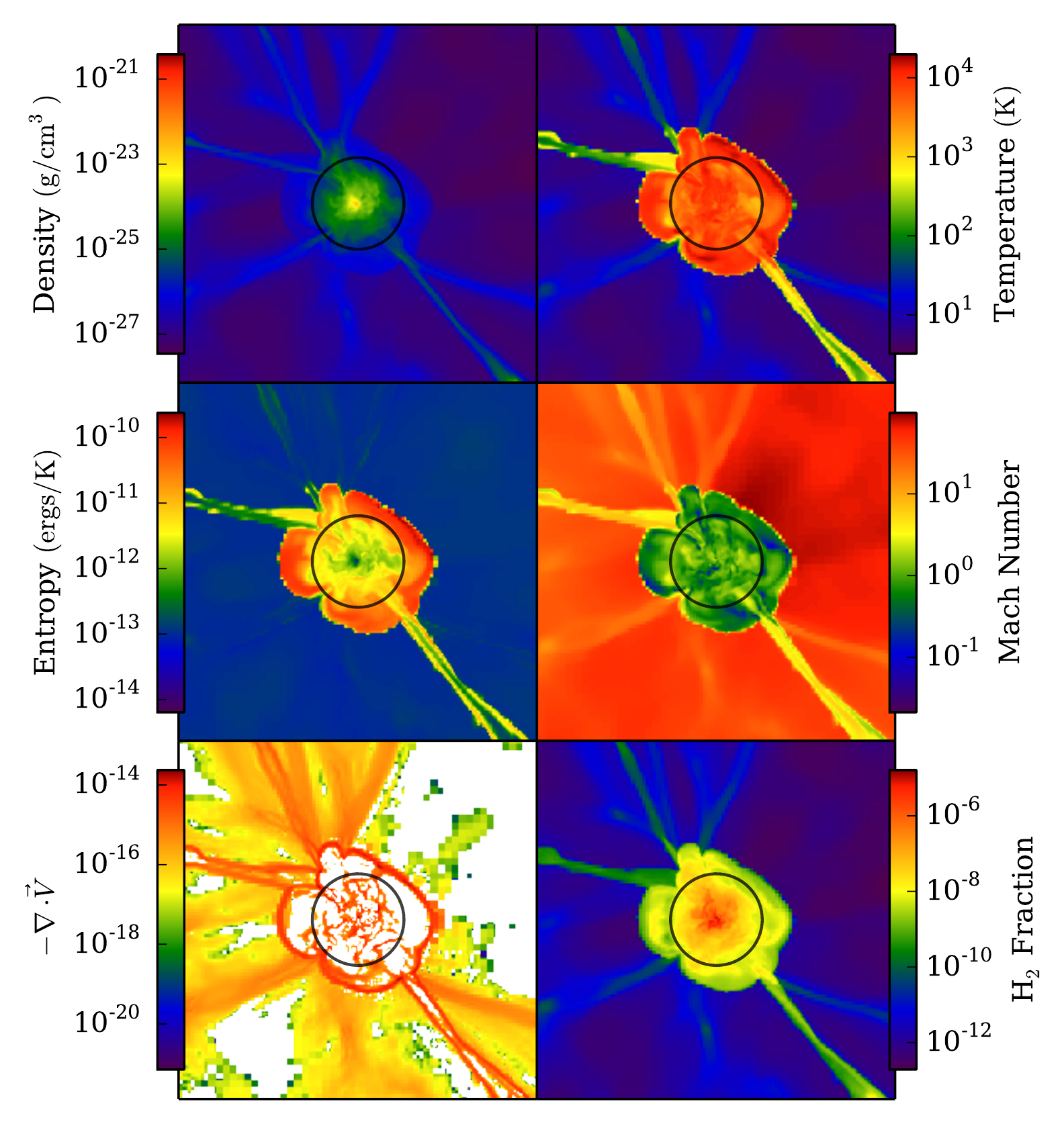} 
\end{center}
\caption{Slices of the y-z plane for density, temperature, entropy, Mach number, $-\nabla \cdot \vec{V}$ and H$_2$ fraction for  halo C (with $T_{\mathrm{vir}}=8.1 \times 10^3$ K) with a low LW background. The field of view of each plot is 5 kpc (physical) and the virial radius is 0.64 kpc.}
\label{slice-plots}
\end{figure*}

The addition of radiative cooling causes the halos to cool and collapse before the redshifts at which they were identified in the low-resolution run.  This occurs because the halos build up in a hierarchical fashion and so have progenitor halos with virial temperatures that allow them to cool via H$_2$ cooling.  We stop the runs when they reach the maximum refinement level allowed in the simulation, as at this point they will start to form stars.   The collapse redshifts, virial masses and virial temperatures of the halos at the point at which they collapse are shown in Table~\ref{high-res-run}.  Only the third halo reaches a virial temperature close to $10^4$ K.  In this context, we define collapse as the point when we would need additional refinement in order to continue evolving the halo, which corresponds to a density of approximately $10^8$ cm$^{-3}$.

To investigate if the density-dependent H$_2$ suppression mechanism is operating, we analyze the simulation results at the point at which they collapse.  In Figure~\ref{phase-plots} we plot the temperature and density distribution of all cells within a radius of a few times the virial radius for each simulation run. Additionally, we overlay the ``zone of no return'' in this figure, using the approximate curves for the boundary of the zone from \cite{Inayoshi2012}. From the plots we see directly that no cells have been shocked-heated into this ``no return zone''. This indicates that the H$_2$ suppression mechanism is not operating, at least for these halos at this point.

\citet{Inayoshi2012} argued that cold, filamentary flows would penetrate to the center of these halos, shock heating the gas at high densities.  To better understand why this is not happening, in Figure~\ref{slice-plots} we plot slices through the center of halo C (with $T_{\mathrm{vir}}=8.1 \times 10^3$ K), showing the density, temperature, entropy, negative of the divergence of the velocity field, and H$_2$ fraction. From the density and temperature slices, we clearly see the cold filamentary structure feeding the halo. However, it is evident that no cold flows (T $\sim 10^{2} - 10^{3}$ K) penetrate unperturbed into the center.  Instead, the gas quickly heats to the average temperature around the virial radius. This is consistent with the phase plots which show continual heating of the cells at number densities approach $\sim 1$ cm$^{-3}$. Most of the cells at this number density have roughly T $\sim 10^4$ K. At higher densities $\gtrsim 1$ cm$^{-3}$, H$_2$ cooling becomes dominant, leaving the gas roughly at $\sim 700$~K.

One way to identify shocks is to look for sources of entropy production (although shocks are not the only source of entropy).  Figure~\ref{slice-plots} shows that the largest entropy production happens around the virial radius. This is largely due to the spherical accretion of the surrounding cold gas. This is more apparent in the Mach number slice, where the flow outside the virial radius is supersonic and the flow interior to the virial radius is mostly transonic. The transition from supersonic to subsonic flow happens abruptly. We also see the cold inflows are not immune to this shock, and gas in the filaments begins to exhibit higher entropy at the virial radius (although the entropy is considerably lower than most of the gas at the virial temperature). The entropy generation is at the expense of its kinetic energy as evident in the Mach number slice where the inflow also becomes transonic as it passes the virial radius.  The negative divergence of the velocity shown in Figure~\ref{slice-plots} reinforces our previous statements. Shocks produce a large value of $-\nabla \cdot \vec{V}$. It is evident there is a strong shock at the virial radius and the filamentary structure is disrupted as it passes through this radius.  

This shows clearly that the cold, filamentary flows in these halos shock at or around the virial radius, where their densities are low -- too low for the H$_2$ suppression mechanism suggested in \citet{Inayoshi2012} to operate.  Instead, the shocked filaments can efficiently form H$_2$ and cool.

We argue that this conclusion is largely consistent with (comparable) previous work.  For example, \citet{WiseAbel2007} carried out similar simulations, finding that although filamentary gas can, in some cases, penetrate through the virial radius without shocking, it does not get past one-third of the virial radius (as stated in the abstract of that paper). Moreover, this occurs only in the absence of H$_2$ cooling (i.e. in their H+He only runs A6 and B6), which is consistent with our own findings.  In any case, this penetration is not nearly far enough -- for example, Fig. 2 of that paper demonstrates that the densities do not get to $10^4$ cm$^{-3}$ until about 0.005 $r_{\rm vir}$.  The SPH simulations of \citet{Greif2008} also examine the inflow of filaments in halos of similar size.  They appear to find that cold filaments may not shock-heat as the flow in; however two points are relevant here.  The first is that later work \citep{Nelson2013} has shown that SPH simulations incorrectly predict substantial amount of cold-accretion compared to Eulerian or moving-mesh codes.  The second point is that even in the simulations of \citet{Greif2008}, gas does not get close to the H$_2$ zone of no return (e.g. their Fig. 7).  Therefore, we conclude that our key results are in agreement with previous work.

We have seen that -- for the three halos simulated -- the halos collapse and fragment without any gas entering the ``zone of no return''.  They do this in part because the $10^4$ K halos do not collapse monolithically -- instead, the hierarchical formation naturally involves lower temperature halos which can cool efficiently via H$_2$.  

One possible objection is that we have simply not simulated enough halos, and that a small fraction of halos might collapse quickly enough for the suppression mechanism to be efficient.  We examine this possibility in more detail in Section~\ref{sec:discussion}; however, first (in the next section), we look at what would happen if we could somehow suppress cooling in the lower mass halos until high-$T_{\rm vir}$ halos build up.

\begin{figure}
\begin{center}
\includegraphics[scale=0.44]{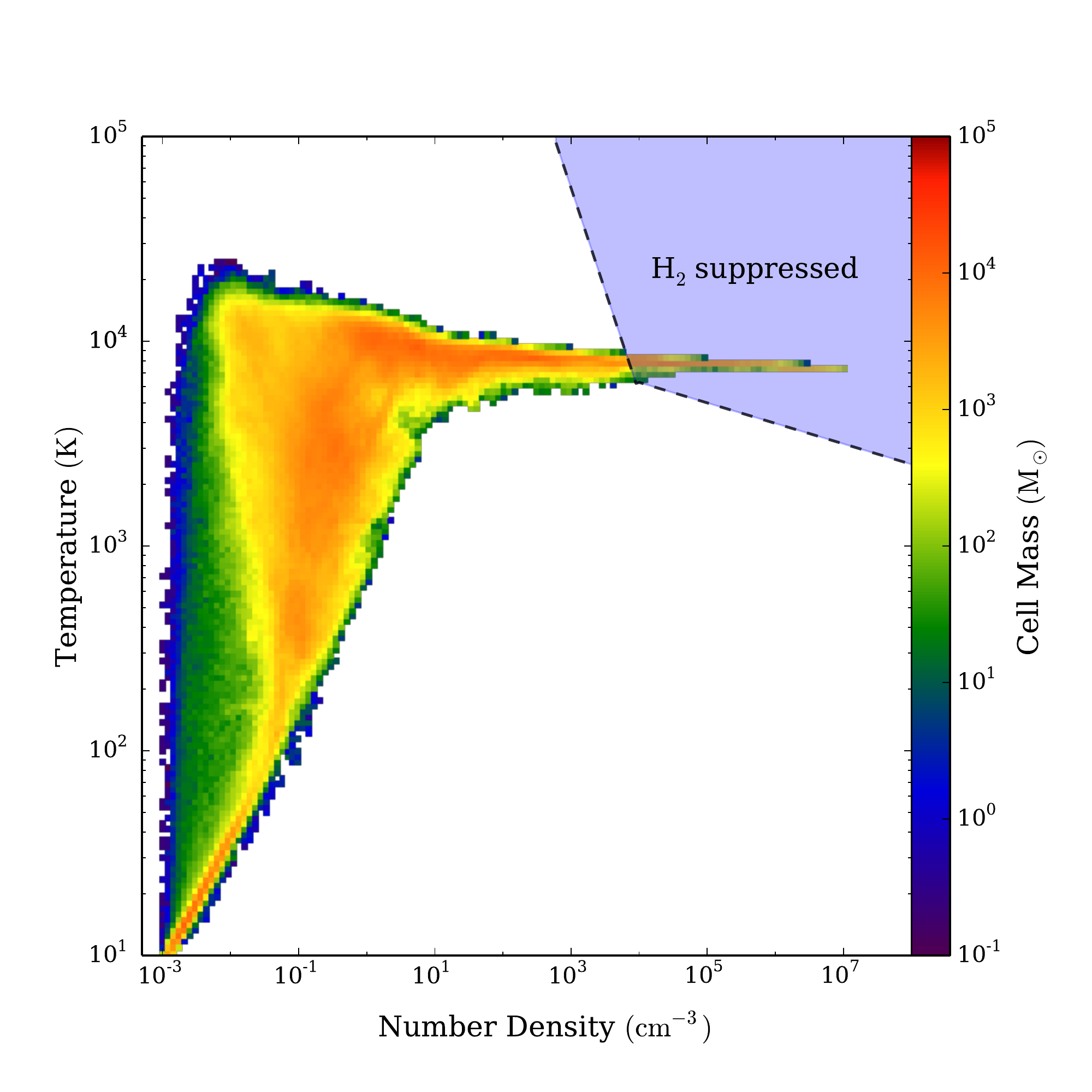} 
\end{center}
\caption{Phase plot of the number density and temperature (with color showing the gas mass distribution) for a simulation of halo A with a large UV background ($J_{21} = 10^5$). The high flux suppresses cooling and we see a significant number of cells in the ``zone of no return''.}
\label{phase-plots-high-J}
\end{figure}

\begin{figure*}
\begin{center}
\includegraphics[scale=0.9]{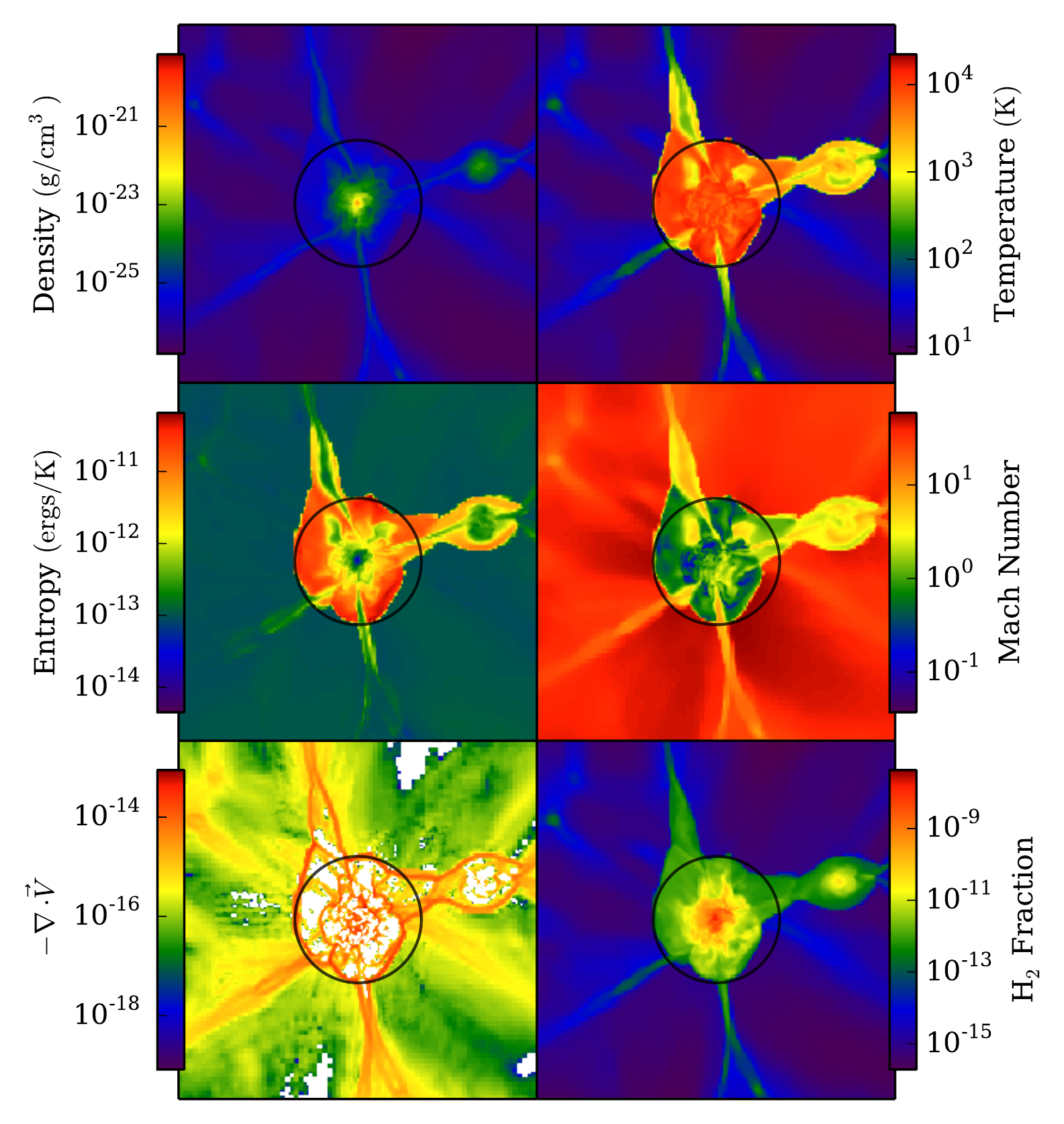} 
\end{center}
\caption{Slices of the y-z plane for density, temperature, Entropy, Mach number, $-\nabla \cdot \vec{V}$ and H$_2$ fraction  for Halo A with a high-UV background ($J_{21} = 10^5$). The field of view of each plot is 3 kpc and the virial radius is 0.53 kpc.}
\label{slice-plots-high-J}
\end{figure*}

Before examining that issue, we make a brief remark about resolution -- a concern with any result from a numerical simulation.  Although we have not explicitly carried out a resolution study as part of this work, this point has been examined regularly in the past.  For example, \citet{Machacek+2001} explicitly ran multiple simulations (also with the Enzo code) to examine how mass resolution impacted the cooling of primordial halos like the ones simulated here.  They used dark matter particles masses of 306 and 38 \msun, finding only minor changes in the statistics of cooling halos, implying that, with a dark matter particle mass of 85 \msun, our results are robust to mass resolution.  A related but slightly different concern is the resolution of the Jeans length -- recent work has suggested that the Jeans length may, in some circumstances, need to be resolved by more than the 16 cells adopted here.  For example, \citet{Turk2012} find that the magnetic properties require ratios greater than 32, with possible changes to the temperature profile as the resolution is changed.  However, the temperature changes are not found in \citet{Greif2013}, and moreover, the impact is only significant at higher densities than explored here, $n \gtrsim 10^{12}$ cm$^{-3}$.  At lower densities, the profiles are well converged.  Therefore, we conclude that this issue is unlikely to impact the results found here.
We can also compare our profiles to previous work.  For example, \citet{Oshea2008} carried out simulations of similar halos with similar LW backgrounds -- the temperature profiles they find (e.g. their Fig. 9) are very similar to ours.  As that paper demonstrates, the relatively high temperatures ($\sim 700$ K) we find in the H$_2$ cooling region are an indirect consequence of the relatively high LW background used here.


\subsection{$J _{21} = 10^5$ Run}

To investigate if the collisional H$_2$ suppression mechanism can keep a halo from fragmenting if we could place its central region in or near the ``zone of no return'', we re-simulate halo A using a much higher value of the LW-background to artificially suppress H$_2$ cooling.  We adopt $J_{21} = 10^5$ -- with such a large (and unrealistic) background, most of the H$_2$ will be dissociated, as shown by \cite{Shang2010} and \cite{Wolcott-Green2011}, where the critical flux was determined to be $J_{21} = 10^3$. We picked the halo with the smallest virial temperature at collapse in the previous runs ($T_{\mathrm{vir}}=2.86 \times 10^3$ K) and reran it with the large $J_{21}$. The halo collapses at a later redshift, $z=16.74$, with a larger virial temperature $T_{\rm vir} = 0.97 \times 10^4$ K, mass $2.48 \times 10^7$ M$_{\odot}$, and radius $534$ pc.

\begin{figure*}
\begin{center}
\includegraphics[scale=0.8]{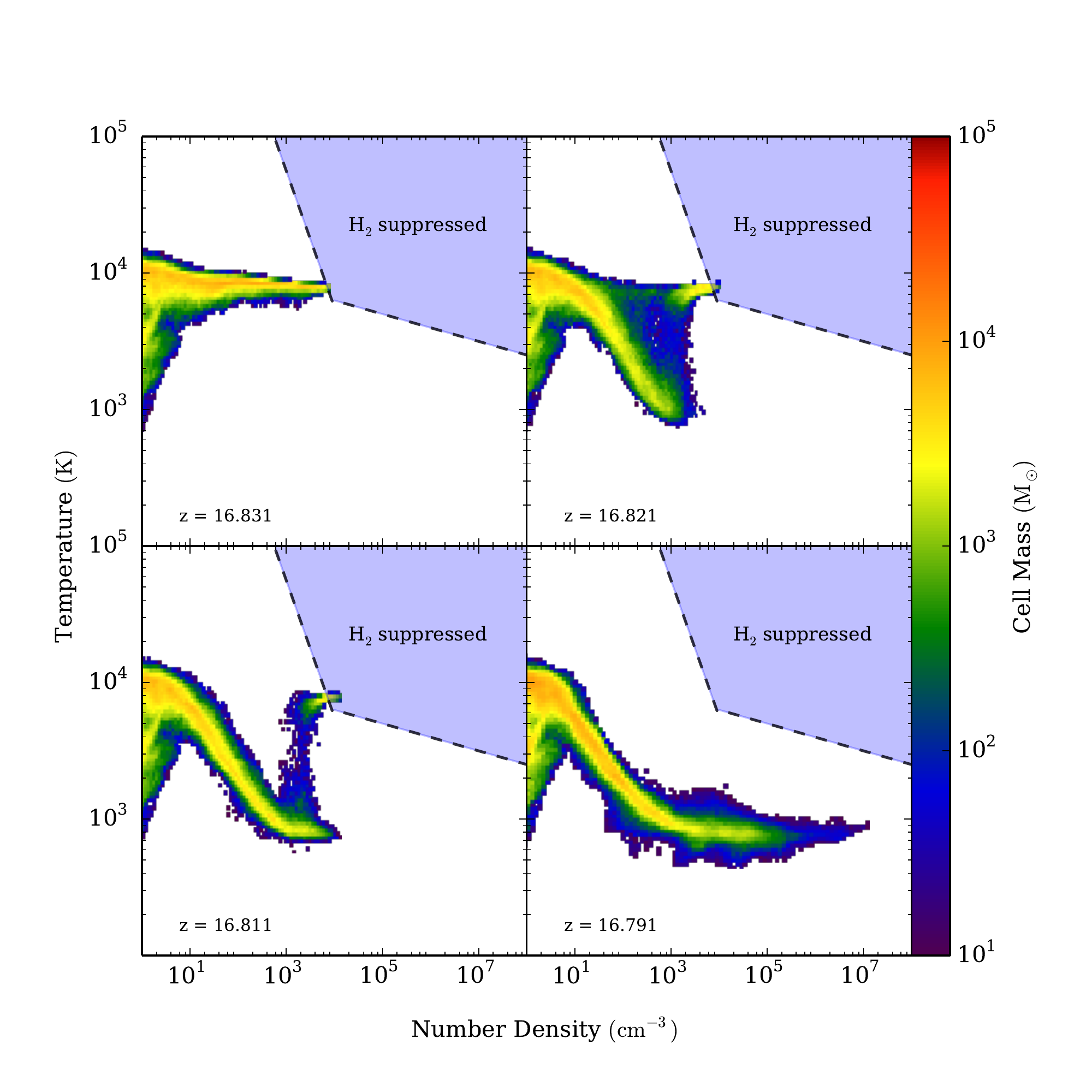} 
\end{center}
\caption{Phase plots showing the distribution of number density and temperature in halo A at a variety of redshifts, as indicated. The simulation ran with $J_{21} = 10^5$ until the maximum density reached the edge of the ``zone of no return'' (as shown in the top left panel). Thereafter the gas is evolved with a reduced $J_{21} = 10$.  Gas quickly forms H$_2$ and cools to a temperature well below the zone.}
\label{phase-plot-high-J-one}
\end{figure*}

\begin{figure*}
\begin{center}
\includegraphics[scale=0.46]{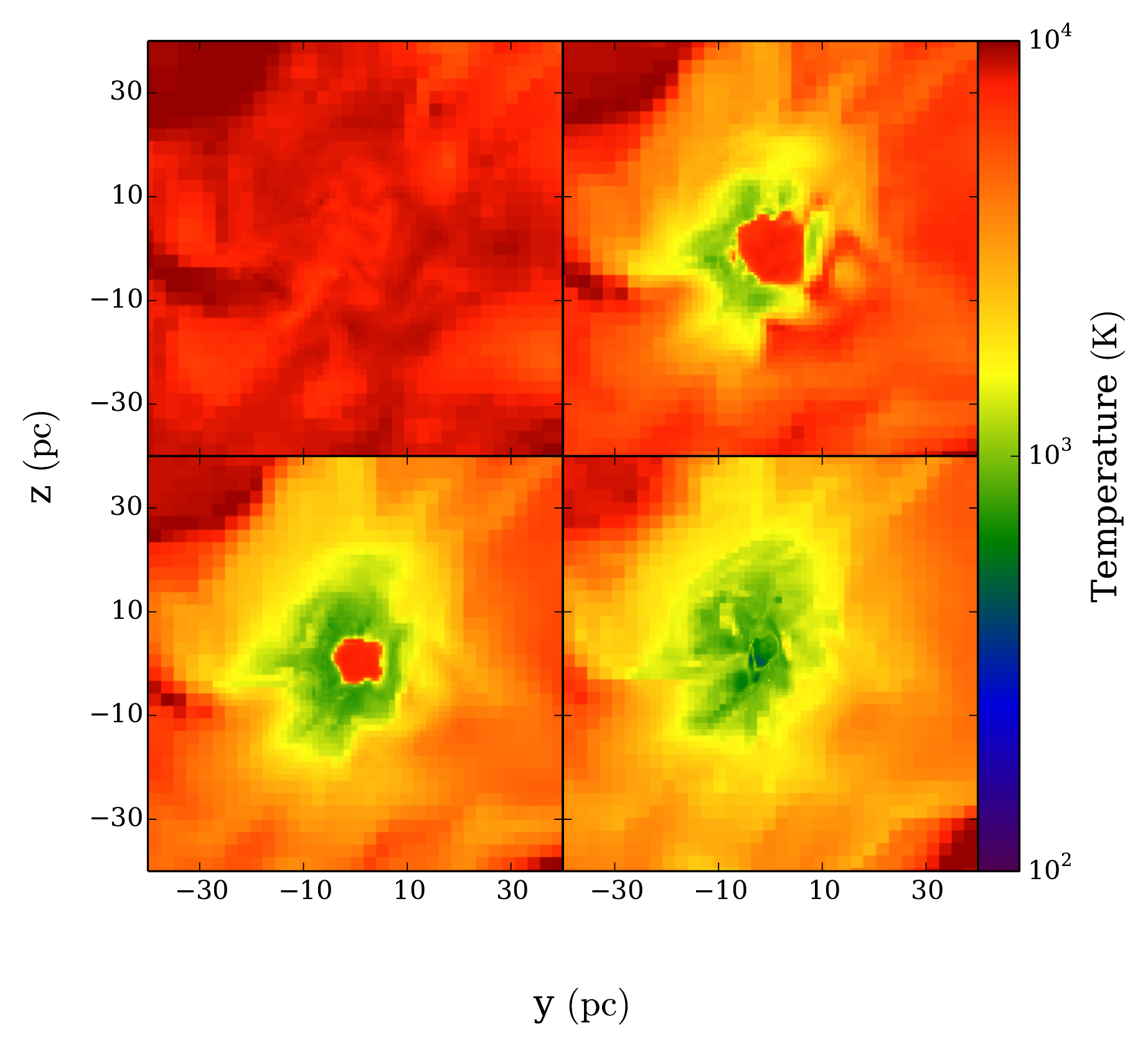} 
\includegraphics[scale=0.46]{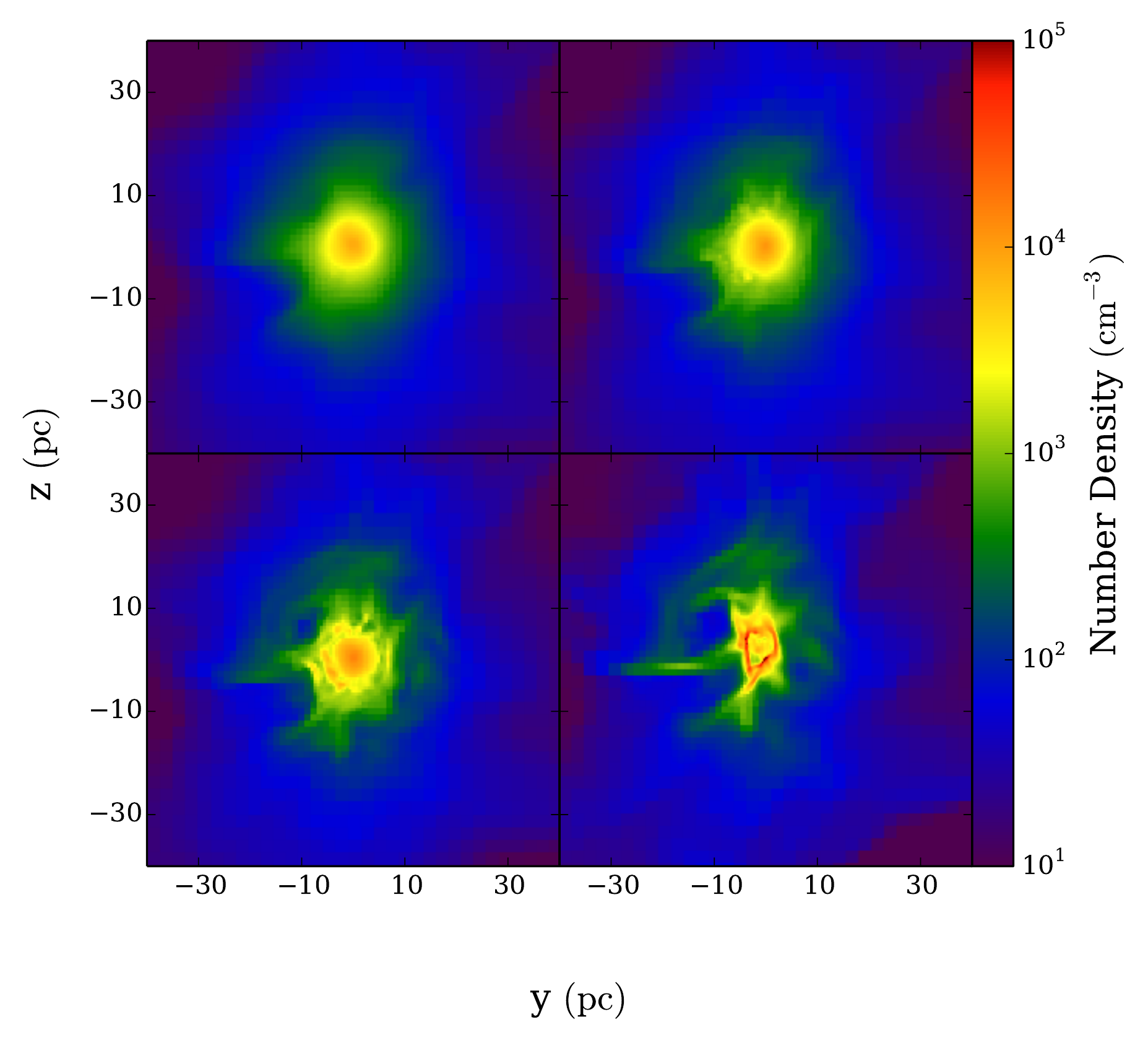} 
\end{center}
\caption{The left set of four panels shows temperature slices through the center of halo A corresponding to the same simulation and output times as in Figure~\ref{phase-plot-high-J-one}, while the right set of four panels shows number density slices for the same halo at the same sets of times.  After the background flux is reduced, the lower-density region outside the core cools first, driving an outflow which evaporates the center.}
\label{slices-high-J-one}
\end{figure*}

In Figure~\ref{phase-plots-high-J}, we show the number density and temperature phase plot for this run. We see the cells for $n \gtrsim 1$ cm$^{-3}$ have now increased in temperature. The large $J_{21}$ has dissociated most of the H$_2$, leaving atomic cooling as the dominant mechanism and keeping the gas at $T\sim 8000$ K. For $n \lesssim 1$ cm$^{-3}$ the same structure is apparent as the previous simulations, where cells are heated to $T\sim 10^4$ K at the virial radius.

Figure~\ref{slice-plots-high-J} shows slices with the same quantities as in the previous section, but now for the large $J_{21}$ run. Although the general structure is quite similar, there are now some signs of cold flows penetrating past the virial radius. This is more apparent in the Mach number slice where we see high velocity flows passing the virial radius where the majority of the fluid is transonic.  It can also be seen in the temperature distribution, where low-temperature gas flows into the halo before heating. However, we note that the ``zone of no return'' occurs at densities above $\sim 10^4$ cm$^{-3}$, which corresponds to the very central region (colored red) of the density slice, and so apparently, the cold flows are not penetrating this zone.


\subsubsection{Stability of the Zone of no Return}

Next, we investigate if gas in the ``zone of no return'' will indeed stay there without any external flux.  We do this by turning down the high-UV background flux in the previous simulation. We run two identical simulations as in the last section, but now reduce $J_{21}$ at different evolutionary points. The two points are distinguished by whether or not gas cells have reached the critical H$_2$ number density for local thermodynamic equilibrium, $n_{\mathrm{cr}}\sim 10^4$ $\mathrm{cm}^{-3}$. Thus, in our first simulation, we allow the gas to evolve with a large $J_{21}$ until the maximum density just reaches $n_{\mathrm{cr}}$, and at this point the UV background is reduced to $J_{21} = 10$ and the gas is allowed to evolve. Similarly, our second simulation is identical with the exception that the reduction is applied when a small fraction of cells have entered into the ``zone of no return''. 

Figure~\ref{phase-plot-high-J-one} shows the phase plot evolution for the first simulation (the instant of flux reduction corresponds to the upper-left panel, at $z=16.831$). As the simulation evolves with the reduced $J_{21}$, H$_2$ begins to form and cool efficiently.  The cooling occurs first at densities $n \sim 10^3 \, \mathrm{cm}^{-3}$, leaving the higher density regions still hot ($T \sim 10^4$ K).  As the evolution progresses, this gas eventually cools and by the final time shown, when the halo collapses to the highest refinement level we allow, the phase distribution looks very much like the low-J runs.

It is interesting to analyze this evolution in more detail.  In Figure~\ref{slices-high-J-one}, we show temperature (left) and density (right) slices that go through the center of this halo for the same times presented in Figure~\ref{phase-plot-high-J-one}.  We see that the cooling process creates a cold envelope surrounding the hot core.  This core is over-pressured, which generates an outflow, driving clumps of dense material out of the central region, as evident in the density slices of Figure~\ref{slices-high-J-one}.  The outflow reduces the density (and temperature) of the gas in the core, which is then effectively evaporated.  The final remnant is a dense shell surrounding a core that has cooled on average to $T\sim 700$~K.

\begin{figure*}
\begin{center}
\includegraphics[scale=0.8]{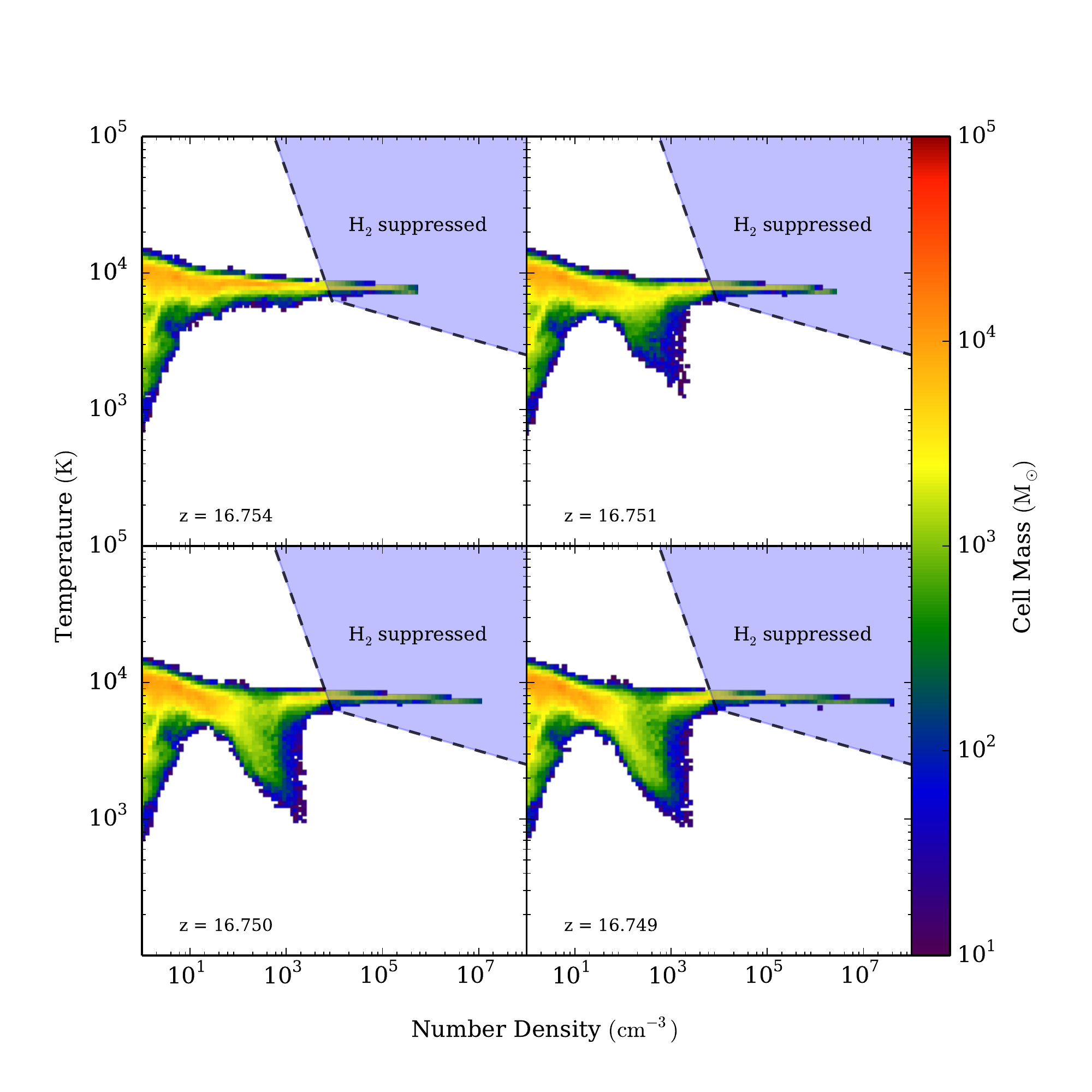} 
\end{center}
\caption{Phase plots showing the distribution of number density and temperature in halo A at a variety of redshifts, as indicated. The simulation ran with $J_{21} = 10^5$ until the maximum density reached well into the ``zone of no return'' (top left panel). Thereafter the gas is evolved with a reduced $J_{21} = 10$.  Although some gas at moderate densities begins to cool, the gas in the ``zone of no return'' collapses more quickly than the outer region can cool.}
\label{phase-plot-high-J-two}
\end{figure*}

In the second simulation of this section, we allow the run with $J_{21} = 10^5$ to continue a bit further, to $z=16.754$, when a significant amount of gas has entered the ``zone of no return'' (see the top left panel of Figure~\ref{phase-plot-high-J-two}) before turning the UV background down to $J_{21} = 10$.  The resulting evolution of the phase diagram is shown in Figure~\ref{phase-plot-high-J-two}. In this case the evolution is quite different: although some gas does cool, the dense gas evolves more quickly, and by the time we stop the simulation (when it reaches our highest refinement level at $z = 16.749$), the amount of mass inside the ``zone of no return'' has not changed by more than a few percent. There is some cooling -- in particular at $n \sim 10^3$ cm$^{-3}$ and $n \sim 10^6$ cm$^{-3}$; however, this does not significantly affect the evolution of the clump. 
 
The resulting baryonic mass in the ``zone of no return'' is $\sim 10^5$ M$_{\odot}$, roughly 10\% of the halo mass within the virial radius.

 
\section{Discussion}
\label{sec:discussion}

Using a set of numerical simulations, we have investigated the viability of the mechanism for collisional suppression of H$_2$ suggested by \citet{Inayoshi2012}.  We found that, for the three halos examined, H$_2$-mediated cooling set in before the conditions required for suppression could be established -- in particular, before a $T_{\rm vir} \gtrsim 10^4$ K halo forms, which would be a pre-requisite for the generation of shocks strong enough to put gas into the ``zone of no return''.  

However, we did confirm (using a set of simulations that employed a high UV background to temporary suppress cooling) some aspects of the mechanism: if the gas in the core of an atomic cooling
halo can avoid ${\rm H_2}$ cooling until it enters the ``zone of no return'', then
subsequent ${\rm H_2}$ cooling can be naturally averted. The gas would then
remain at $\sim 10^4$K, possibly leading to conditions suitable for
SMBH formation, as envisioned in the many previous works mentioned in
the Introduction (see, e.g. \citealt{Haiman2013} for a recent review).

Unfortunately, H2 cooling at any time in the past history of the
same gas would likely invalidate this possibility, if this H2 cooling
allowed the gas to reach temperatures of a few 100 K and to evolve to
high density at such low temperatures. This cold and dense gas, at
the center of the progenitor minihalo, would form one, or possibly a
few, PopIII stars. The subsequent evolution of this progenitor into
an atomic cooling halo is uncertain, however, there are many obstacles
to rapid SMBH formation.

First, even a handful of SNe (or a single pair instability SNe) can enrich
the entire atomic cooling halo to a metallicity of $Z \gtrsim 10^{-3} {\rm Z_\odot}$
\citep[e.g.,][see page 396]{BrommYoshida2011} the critical value above which
direct SMBH formation is replaced by fragmentation \citep{Inayoshi2012}.
It is possible that none of these Pop III stars produce any metals
\cite[such as non-rotating metal-free stars between 40-140 \msun;][]{Heger2003}.
Such massive Pop III stars would leave behind stellar-mass
BHs, which, in principle, could grow rapidly into SMBHs, provided that
they are surrounded by very dense gas, and accrete sufficiently
rapidly to trap their own radiation \citep[e.g.,][]{VolonteriRees2005}.
However, the parent Pop III stars of these seed BHs will create a
large ionized bubble and therefore the seed BHs will likely begin
their life in a low-density medium. As they later begin to accrete,
their radiation likely self-limits the time-averaged accretion rate to
a fraction of the Eddington rate \citep{Alvarez2009, ParkRicotti2012, 
Milosavljevic2009}.  Although the above scenarios are worth investigating further in
detail, we will hereafter assume that these obstacles prevent rapid
SMBH formation in any halo that experienced H$_2$ cooling at any time in
its history.

\subsection{Avoiding ${\rm H_2}$ Cooling by Rapid Halo Assembly}

However, since we only simulated three halos a natural
question is whether ${\rm H_2}$ cooling may be avoided in a few,
highly atypical, atomic cooling halos, even if the background flux
$J_{\rm LW}$ was much lower.  One possibility is that -- in rare cases
-- the progenitor halos could experience unusually rapid mergers,
continuously shock-heating the gas on a time-scale that always remains
shorter than the ${\rm H_2}$--cooling time.  Here we evaluate the
likelihood of this scenario, using Monte Carlo realizations of the
merger histories of $10^6$ atomic cooling halos.

In particular, we start by creating dark matter halo merger trees,
using the Monte Carlo algorithm in \citet{Zhang+2008b}.  This paper
presents three different numerical algorithms, which are based on the
conditional halo mass functions in the ellipsoidal collapse model
\citep{ShethTormen2002,Zhang+2008a}.  We adopted their ``method B'',
which was found in subsequent work to provide the best match to the
statistics of merger trees in N-body simulations (Zhang 2012, private
communication). This method represents an improvement over previous
merger-tree methods based on spherical collapse -- in particular, the
ellipsoidal collapse models tend to predict a larger number of more
massive progenitors, or a ``flatter'' assembly history
\citep{Tanaka2013}.

We have created $10^6$ merger trees of a DM halo with $T_{\rm
vir,0}=10^4$ K ($M_0=6.3\times10^7~{\rm M_\odot}$) at redshift
$z_0=10$, extending back to redshift $z=20$.  We then follow the mass
of the most massive progenitor back in time, and at each redshift $z$,
we compute the ${\rm H_2}$--cooling time 
\begin{equation}
t_{\rm H2} = {1.5 n_g k_B T \over \Lambda n_H n_{H2}}\end{equation} 
in this progenitor, using the procedures and rates in \citet{Machacek+2001}.
Specifically, we compute the maximum density 
\begin{equation}
n_{\rm max} = 187 \Omega_b h^2 (T_{\rm vir}/1000 K)^{1.5} {\rm cm}^{-3}
\end{equation} 
that could be reached by the gas in the nucleus of this progenitor, in the
absence of ${\rm H_2}$--cooling (i.e. by adiabatic compression).  We
then specify a constant background flux $J_{\rm 21}$, and compute the
equilibrium ${\rm H_2}$ fraction and the corresponding $t_{\rm H2}$ at
this density, given $J_{\rm 21}$ and $\rho_{\rm max}$.  Finally, we
require that the {\em longer} of the dynamical time $t_{\rm
dyn}=\sqrt{3\pi/16G\rho_{\rm max}}$ and $t_{\rm H2}$ remains longer
than the cosmic time $\Delta t(z) = t(z_0)-t(z)$ elapsed between the
$z$ and $z_0$.  This last requirement ensures that the most massive
progenitor of our atomic cooling halo at redshift $z>z_0$ cannot cool
via ${\rm H_2}$ {\em and} evolve dynamically, before it is
incorporated by mergers into the $M_0=6.3\times10^7~{\rm M_\odot}$
atomic cooling halo at redshift $z_0=10$.  Note that this requirement
must be satisfied at {\em all} redshifts $z>z_0$ -- in other words, if
our atomic cooling halo had a progenitor, at any redshift $z$, that
was massive enough to cool and collapse, the ``SMBH formation by
direct collapse'' scenario is no longer feasible.

In practice, as we run our merger trees backward, we simply look, at
each redshift step, whether the most massive progenitor violates the
criterion $\Delta t(z)< {\rm max}[t_{\rm H2},t_{\rm dyn}]$. If it
does, we record the redshift $z_{\rm max}$ where this happened, and we
discard the rest of this merger tree.

\begin{figure}
\centering
\mbox{\includegraphics[width=8cm]{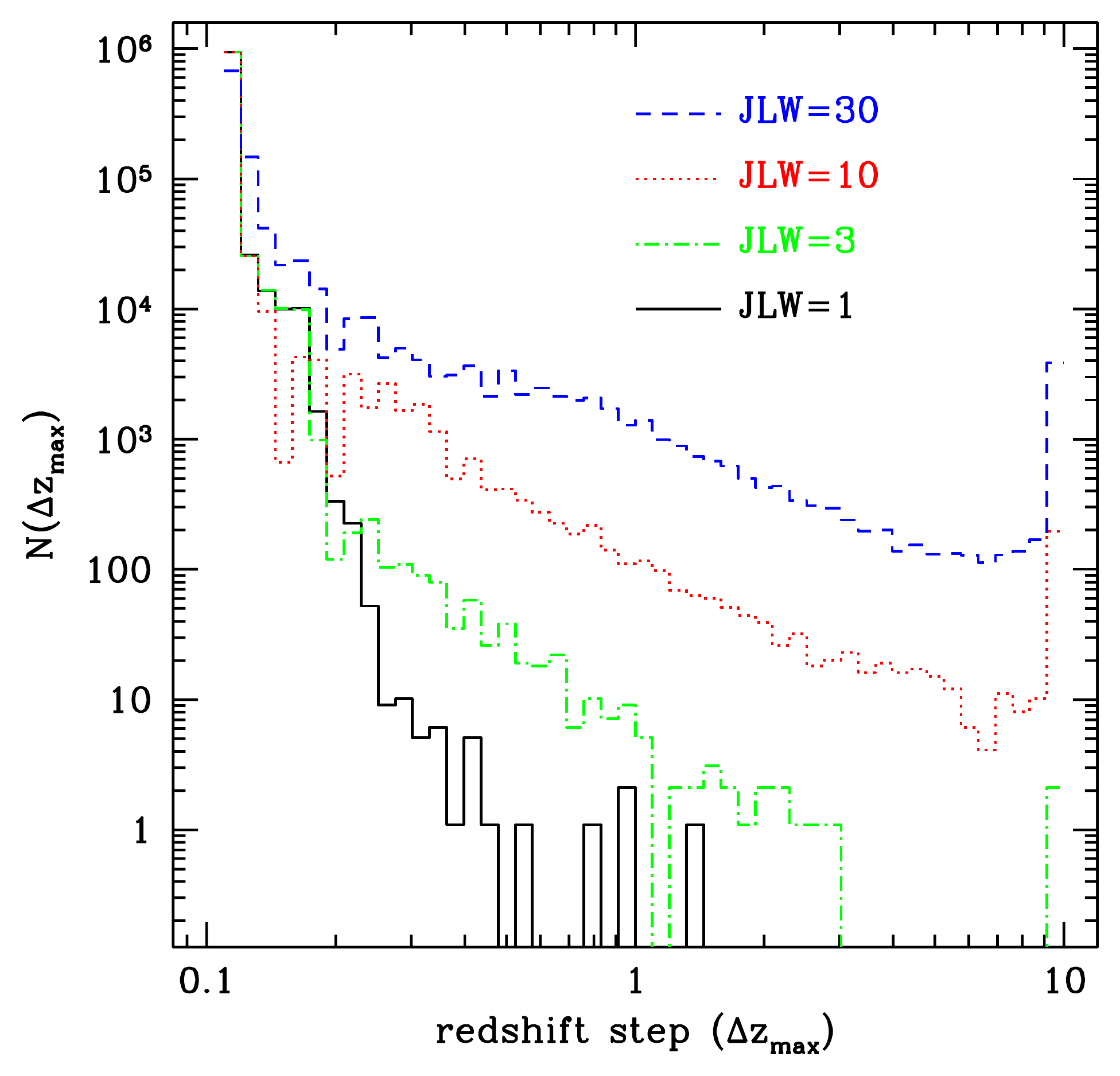}}
\caption{\label{fig:zmax} The figure is derived from $10^6$ Monte
Carlo realizations of the merger tree of an atomic cooling halo with
$T_{\rm vir,0}=10^4$K ($M_0=6.3\times10^7~{\rm M_\odot}$) at redshift
$z_0=10$.  It shows the probability distribution of the terminal
lookback redshift $\Delta z_{\rm max}$, at which the atomic cooling
halo had a progenitor, which was massive enough to cool via ${\rm
H_2}$ and collapse.  The ``SMBH formation by direct collapse''
scenario is feasible only if none of the progenitors can cool and
collapse prior to $z_0=10$.  For a LW background $J_{\rm LW}=1$ (shown
by the solid black histogram), all of the $10^6$ merger trees contain
an ${\rm H_2}$--cooling progenitor.  However, for higher background
fluxes of $J_{\rm LW}=3$ (in dot-dashed green), $J_{\rm LW}=10$ (in
dotted red), or $J_{\rm LW}=30$ (in dashed blue), we find
approximately 2, 200, and 4000 cases, respectively, where none of the
progenitors, out to $z=20$, could cool via ${\rm H_2}$ - this is shown
by the pile-up of the probability distribution in the last bin at
$\Delta z_{\rm max}=10$.}
\end{figure}

In Figure~\ref{fig:zmax}, we show the probability distribution of the
``terminal lookback redshift'' $z_{\rm max}$ we have found among the
$10^6$ merger trees, for four different values of a constant $J_{\rm
LW}=1, 3, 10$, and $30$. For a LW background $J_{\rm LW}=1$ (shown in
the figure as the solid black histogram), we found that all of the
$10^6$ merger trees contain an ${\rm H_2}$--cooling progenitor by
$z<20$.  However, for the higher background fluxes of $J_{\rm LW}=3$
(dot-dashed green), $J_{\rm LW}=10$ (dotted red), or $J_{\rm LW}=30$
(dashed blue), we have found approximately 2, 200, and 4000 cases,
respectively, where none of the progenitors, out to $z=20$, could cool
via ${\rm H_2}$ - this is shown by the pile-up of the probability
distribution in the last bin at $\Delta z_{\rm max}=10$.  For the
atomic cooling halos with these particular merging histories, we
expect that ${\rm H_2}$ cooling is avoided entirely until the atomic
cooling halo forms at $z_0=10$.  These rare halos are good candidates
where the subsequent ionizing shocks can prevent the gas temperature
from falling below $\approx 10^4$K, allowing the gas to collapse
directly to a SMBH.  This scenario reduces the required LW flux by a
factor of $\sim 300$, from $J_{21}\approx 10^3$ \citep{Shang2010,
Wolcott-Green2011} to $J_{21} \approx 3$, but it does require that the
latter LW background is in place at a redshift as high as $z\sim 20$
(in order for a fraction $2\times10^{-6}$ of the atomic cooling halos
at $z_0=10$ to be such candidates).

\section{Summary}

One way to explain the presence of SMBHs at $z > 6$ with masses
greater than $10^9$ \msun\ is through the formation of a massive
($\sim 10^5$ \msun) BH seed from gas in $T_{\rm vir} \gtrsim 10^4$ K
halos, either via direct collapse to a BH or through the formation of
a supermassive star or a quasi-star \citep{Begelman2008,
Hosokawa2013}.  As noted in the Introduction, one difficulty with
these modes is the need to prevent fragmentation of the cooling halo
into lower mass stars, and, in particular, to prevent cooling due to
H$_2$.  \citet{Inayoshi2012} have suggested that this can be
accomplished through the action of cold flows, which result in gas
shocking to high densities ($n > 10^4$ cm$^{-3}$) and temperatures ($T
> 10^4$ K).  This shocked gas cools by Ly$\alpha$ emission to about
8000 K; however, if the density is high enough, enhanced H$_2$
collisional dissociation suppresses the gas from cooling further.
This was demonstrated using one-zone models calculated by
\citet{Inayoshi2012}, who identified a ``zone of no return''.  This
mechanism is appealing as it does not require a high LW background to
destroy the H$_2$; however, it is unclear if this mechanism operates
in nature.

To test this idea, we carried out cosmological hydrodynamic simulations with the adaptive mesh refinement code Enzo.  We first identified three halos from a low resolution simulation which all had $T_{\rm vir} \gtrsim 10^4$ K at redshifts ranging from 12 to 17.  We then re-simulated these halos at high-resolution with a relatively modest UV flux in the Lyman-Werner band, $J_{21} = 10$.  We found that in all three cases, cooling from H$_2$ was able to efficiently lower the gas temperature below that of the ``zone of no return'', indicating that, at least for these three halos, the mechanism was not operating.

To determine why this occurred, we examined the structure of the simulated halos in detail, and found that, while cold flows do occur, they generally shock at or near the virial radius and do not penetrate into the halo center where the densities are high.  We note that although these small halos have low virial temperatures and so might be naively classified as ``cold-mode" halos \citep{Birnboim2003, Keres2005, Dekel2006}, they actually have low cooling rates because of the inefficiency of H$_2$ cooling.  Therefore, the characteristic cooling times of these halos is longer (or comparable) to their dynamical times, making them more akin to hot-mode halos.  This is consistent with the clear virial shocks that we see in these halos (e.g., Figure~\ref{slice-plots}).

To determine if the suppression mechanism could function if we artificially suppressed H$_2$ cooling, we reran one of the simulations with a high LW background ($J_{21} = 10^5$), well above the critical flux required to suppress H$_2$ cooling radiatively \citep{Shang2010, Wolcott-Green2011}.  We showed that, in this case, the gas did indeed enter the ``zone of no return''.  To see if this situation was stable, we ran two variations of this simulation, in which we turned the flux off either just before, or just after the gas in the halo center entered the ``zone of no return''.  In the first case, cooling eventually won out, with the region just outside the core cooling first and driving an evaporative wind which led to cooling by H$_2$ throughout the halo.  However, in the second case, the gas inside the ``zone of no return'' stayed there, and eventually collapsed to high densities while remaining at $T \sim 8000$ K, despite the lack of any LW flux.

This demonstrates that the mechanism would work if a halo could collapse quickly enough, but that typical $T_{\rm vir} \gtrsim 10^4$ K halos have progenitors which can cool efficiently.  To investigate if any halos can collapse quickly enough to escape this fate, we ran Monte Carlo merger tree calculations of $10^6$ halos with an ellipsoidal collapse model.  We used a simple analytic prescription to determine if any halos could assemble quickly enough to prevent cooling and collapse prior to building up to a $T_{\rm vir} = 10^4$ K halo.   We found that for a Lyman-Werner background of $J_{21}=1$, none of the $10^6$ merger tree histories collapsed quickly enough. For larger LW backgrounds, a small fraction (up to $\sim 10^{-3}$ for $J_{21}$ as high as 30) of the halos did form quickly enough to evade cooling.  We note that at higher redshift, $J_{21}$ would certainly be expected to be lower.

In summary, we confirm the essential physics proposed by
\citet{Inayoshi2012}, but we find that the supersonic filamentary
flows are unlikely to shock gas into the ``zone of no return'', as
they had originally envisioned. Rather, we propose here a modification
of their scenario. In the core of a few rare halos, which assembled
unusually rapidly, the gas may have been kept continuously
shock--heated throughout their history, and eventually shocked into
the ``zone of no return''.  This scenario reduces the required LW flux
by a factor of $\sim 300$, from $J_{21} \approx 10^3$
\citep{Shang2010, Wolcott-Green2011} to $J_{21} \gtrsim 1$, but it
does require that the latter LW background is in place at a redshift
as high as $z\sim 20$.  Future, self-consistent global models of the
build-up of the LW background, together with the assembly of a large
number of atomic-cooling halos, is required to assess the viability of
this scenario.

\section*{Acknowledgments}

We thank Jun Zhang for useful discussions about ellipsoidal
collapse-based merger trees, and for sharing unpublished results on
comparisons between Monte Carlo methods and N-body simulations.  ZH
acknowledges financial support from NASA through grant NNX11AE05G.
GB acknowledge financial support from NSF grants AST-0908390 and AST-1008134, and NASA grant NNX12AH41G,  as well as computational resources from NSF XSEDE, and Columbia University's Hotfoot cluster.

\bibliography{sms_paper}

\label{lastpage}
\end{document}